\documentclass[aps,prd,notitlepage,nofootinbib,superscriptaddress,
twocolumn,
showpacs]{revtex4-2}
\usepackage{amsmath}
\usepackage{amssymb,amsfonts}
\usepackage{graphicx}
\usepackage{color}
\usepackage{listings}
\usepackage{textcomp}
\usepackage{url}
\usepackage{diagbox}
\usepackage{mathtools}
\usepackage[section]{placeins}
\usepackage{youngtab}
\newcommand{\be}{\begin{equation}}
\newcommand{\ee}{\end{equation}}
\newcommand{\dd}{\mathrm{d}}
\newcommand{\bse}{\begin{subequations}}
\newcommand{\ese}{\end{subequations}}

\newcommand{\ii}{\mathrm{i}}

\newcommand{\bpm}{\begin{pmatrix}}
\newcommand{\epm}{\end{pmatrix}}
\newcommand{\bmm}{\begin{matrix}}
\newcommand{\emm}{\end{matrix}}

\newcommand{\x}{\times}

\allowdisplaybreaks

\makeatletter
\newcommand*{\Relbarfill@}{\arrowfill@\Relbar\Relbar\Relbar}
\newcommand*{\xeq}[2][]{\ext@arrow 0055\Relbarfill@{#1}{#2}}
\makeatother

\usepackage[normalem]{ulem}
\renewcommand\sout{\bgroup\markoverwith{\textcolor{red}{\rule[0.5ex]{4pt}{0.8pt}}}\ULon}

\def\mb{\left(\begin{matrix}}
\def\me{\end{matrix}\right)}

\newcommand{\Psix}[3][1]{
\begin{tikzpicture}[scale=0.8]
\node[name=s, regular polygon, regular polygon sides=6, minimum size=1cm, outer sep=0pt ,draw] at (0,0) {}; %draw the plaquette
\foreach \anchor/\x/\y /\xx/\yy /\b in
{corner 1/0.17/0.17*1.732/-0.11/0.18/1, corner 2/-0.17/0.17*1.732/0.07/0.18/2, corner 3/-0.34/0/-0.15/-0.18/3, corner 4/-0.17/-0.17*1.732/-0.22/-0.05/4, corner 5/0.17/-0.17*1.732/0.2/-0.05/5, corner 6/0.34/0/0.15/-0.18/6}
{
 \draw[shift=(s.\anchor)] (0,0) -- (\x,\y) node at(\xx,\yy) {$#2_{\text{\scalebox{0.7}{$\b$}}}$};
 \ifnum #1=1
 \draw[shift=(s.\anchor),<-,>=stealth', line width=0.01pt] (s.\anchor) -- (\x,\y);
 \fi
 }
\foreach \anchor/\xx/\yy /\a in
{side 1/0/-0.18/1, side 2/-0.18/0.05/2, side 3/0.15/0.05/3, side 4/0/-0.18/4, side 5/-0.18/0.05/5, side 6/0.15/0.05/6}
 \draw[shift=(s.\anchor)]  node at(\xx,\yy) {$#3_{\text{\scalebox{0.7}{$\a$}}}$};
\ifnum #1=1{
  \foreach \anchorr/\anchorf in
   {corner 1/corner 2, corner 2/corner 3, corner 3/corner 4, corner 4/corner 5, corner 5/corner 6, corner 6/corner 1}
   \draw[shift=(s.\anchorr), ->, >=stealth', line width=0.01pt]  (s.\anchorr) -- (s.\anchorf);}
 \else {
  \foreach \anchorb/\anchorw in
   {corner 1/corner 2, corner 3/corner 4, corner 5/corner 6} {
   \node[fill=black, circle, minimum size=0, inner sep=0, outer sep=0, draw] at(s.\anchorb) {};
   \node[fill=white, circle, minimum size=0, inner sep=0, outer sep=0, draw] at(s.\anchorw) {};}
}
\fi
\end{tikzpicture}
}

\usepackage{varioref}

\begin{document}
\title{Semiclassical limit of $SU(3)$ gauge field coherent states: Peakedness and overlap functions}
\author{\ Ye Zhang}  
\affiliation{College of Science, University of Shanghai for Science and Technology, Shanghai 200093, China}
\author{\ Zichang Huang} 
%\authornote{*Corresponding authors}
\email{Corresponding author: zhuang1126@usst.edu.cn} 
\affiliation{College of Science, University of Shanghai for Science and Technology, Shanghai 200093, China}

\begin{abstract}
By using the heat kernel method, we construct diffeomorphism-covariant coherent states for the $SU(3)$ gauge group. We numerically demonstrate that these states exhibit the required semiclassical properties in the semiclassical limit: the peakedness property of the probability distribution and the peakedness property of the overlap function. We also provide the leading order term of the overlap amplitude in the combined limit where $t \rightarrow 0$ and $g\rightarrow g'$.  This work provides the essential tool for deriving effective dynamics for $SU(3)$ gauge fields coupled to gravity via a coherent state path integral.
\end{abstract}

\maketitle

\section{Introduction}
Starting from Schr\"odinger's wave packets of the harmonic oscillator in 1926, coherent states have been developed as a powerful tool to bridge the discrepancy between the quantum waves and classical trajectories\cite{schrodinger,doi:10.1142/0096,PhysRev.131.2766}.
At the kinematical level, coherent states---labeled by points in the classical phase space---are expected to be sharply peaked at the corresponding classical configuration in both configuration and momentum representations. 
At the dynamical level, the saddle-point approximation of the coherent state path integral provides one of the best ways to analyze semiclassical dynamics.

In the context of loop quantum gravity (LQG)\cite{book,book1, rovelli2014covariant,han2007fundamental}, the quest for a consistent semiclassical limit, which recovers the general relativity, remains a central challenge for the theory.
Naturally, the coherent states are the best candidates to fulfill this task.
The general construction of the diffeomorphism-covariant coherent states is proposed by Thiemann in his series work \textit{Gauge Field Theory Coherent States}\cite[etc.]{ThomasThiemann_2001,TThiemann_2001,TThiemann_20013}, and examples are explicitly constructed for gauge group $U(1)$ and $SU(2)$.
To maintain the diffeomorphism covariance, the coherent states are defined on a given graph $\gamma$ embedded in the spatial manifold $\Sigma$. 
The classical phase space is given by the $SU(2)$ connection $A$ and the conjugate frame field $E$, which can be smeared to the holonomies $h_e(A)$ along edges $e$ of $\gamma$ and the fluxes $P^i_e(A,E)$ across surfaces dual to edges $e$.
The complexification of the $SU(2)$ holonomy $h_e(A)$ yields an element $g_e$ in the complexified group $SL(2,\mathbb{C})$, which can be identified as a point in the classical phase space $T^*SU(2)$.
The heat kernel complexifier method, a generalization of the Segal-Bargmann transformation, is used to generate the coherent states labeled by complexified $g_e$ in each edge. 
Such coherent states on each edge are proof to be gauge covariant\cite{HALL1994103, Driver1999, hall1999coherentstatesyangmillstheory}, and their direct product over all edges in the graph $\gamma$ gives the coherent states on the graph $\gamma$.
Thus, the coherent states on the graph $\gamma$ are labeled by a point in the classical phase space $T^*SU(2)^{|E(\gamma)|}$, where $|E(\gamma)|$ is the number of edges in $\gamma$.
The coherent states constructed in this way are shown to have the desired -classical properties for gauge group $U(1)$ and $SU(2)$, including the peakedness property, Ehrenfest property and overlap property.

In \cite{Han:2019vpw}, Thiemann's $SU(2)$ coherent states were used in the coherent state path integral formulation after the gauge fixing procedure through deparametrization, based on the semiclassical expansion in the framework of algebraic quantum gravity \cite{Giesel_2007_1,Giesel_2007_2}.
The effective theory derived from the saddle-point approximation of this path integral yields the corrections in the classical limit in cosmology and black hole settings caused by quantum gravity effects \cite{PhysRevD.102.024083,Han:2020uhb,Han:2020iwk,Han:2021cwb,Zhang:2022vsl,Giesel:2024mps,Ferrero:2025est,99fq-xz2w,Long:2021lmd}.
These results demonstrate the power of the coherent state path integral in analyzing the semiclassical limit of LQG.
However, the $SU(2)$ gauge group is not sufficient to describe all the fundamental interactions in nature.
The standard model of particle physics is based on the gauge group $SU(3)\otimes SU(2)\otimes U(1)$, where the $SU(3)$ gauge group describes the strong interaction.
Therefore, it is necessary to generalize the $SU(2)$ coherent states to the $SU(3)$ case, which represents a fundamental step toward developing a semiclassical framework for non-Abelian Yang-Mills theories within quantum gravity.

This generalization, however, presents significant technical challenges. The $SU(3)$ group is of higher rank, with a more complex representation theory and a larger, 16-dimensional phase space $T^{*}SU(3)$.
This complexity profoundly affects the structure of the heat kernel, the character formula and the analytical proofs of semiclassical properties.
Besides, the matrix exponential functions for $SU(3)$ group elements are more complicated than those for $SU(2)$, making the computations of the overlap functions more challenging.

In this paper, we overcome these challenges, construct the $SU(3)$ coherent states through the heat kernel complexifier method, and show that the $SU(3)$ coherent states also have the desired semiclassical properties, including the peakedness property and overlap property.
This work provides the essential tool for future semiclassical analyses of the strong interaction coupled to gravity in a background-independent framework. The paper is structured as follows: Section II reviews Thiemann's formalism to construct diffeomorphism-covariant coherent states. 
Section III details the dimensions and characters of $SU(3)$ irreducible representations and construct the coherent state through heat kernel complexifier method. 
Sections IV, V, and VI present the semiclassical analysis of the peakedness properties and overlap functions, respectively. 
Section VII discusses physical implications and concludes the whole work.

\section{Diffeomorphism-covariant Coherent States}
We begin with a brief review of the construction of diffeomorphism-covariant coherent states for a general compact gauge group $G$ in \cite{TThiemann_2001}.
The classical phase space of a gauge field theory is coordinatized by a $G$-connection field $A$ and a Lie algebra valued, weight one vector density field $E$ on a three-dimensional spatial manifold $\Sigma$.
Let $Lie(G)$ denote the Lie algebra of $G$, with rank $N-1$. 
The generators $\{\tau_i\}, i=1,2,\cdots, N^2-1$ of $Lie(G)$ are normalized according to $\mathrm{tr}(\tau_i\tau_j)=-N\delta_{ij}$, and satisfy the commutation relation $[\tau_i,\tau_j]=2\ii f_{ij}^k\tau_k$, where $f_{ij}^k$ are the structure constants of $Lie(G)$.
The Poisson bracket between $A$ and $E$ is given by
\be
\{A_a^i(x),E_j^b(y)\}=\kappa\ii\delta_a^b\delta_j^i\delta(x,y).
\ee
where $\kappa$ is the coupling constant of the gauge field theory.

As a background-independent theory of gravity, LQG employs holonomies and fluxes to describe the classical phase space and, thereby, manifestly preserves the diffeomorphism covariance.
To define these, we introduce a graph $\gamma$ embedded in $\Sigma$, consisting of edges $e$ and vertices $v$.
The holonomy of the connection $A$ along an edge $e$, which is parametrized by $s\in [0,1]$, is defined as
\be
h_e(A)=\mathcal{P}\exp\left(\int_0^1  A_a^i(e(s))\tau_i \frac{\dd e^a}{\dd s} ds\right),
\ee
where $\mathcal{P}$ denotes path ordering.
Let $S_e$ be the surface dual to the edge $e$, $p_e$ be the intersection point, and $\rho_e(x)$ be a path within $S_e$ connecting $p_e$ and $x$.
The flux of $E$ across $S_e$ is then given by
\be
\begin{split}
&P_{e}^{i}(A,E) \\
=& -\frac{1}{N}\mathrm{tr}\bigg(\tau_{i}h_{e}(0,1/2) \\
&\times\left(\int_{S_{e}}h_{\rho_{e}(x)}*E(x)h_{\rho_{e}(x)}^{-1}\right) \\
&\times h_{e}(0,1/2)^{-1}\bigg),
\end{split}
\ee
where $\ast E$ is the Hodge dual of $E$ and
$h_e(0,1 / 2)$ is the holonomy along the first half of the edge $e$ (from $s=0$ to $s = 1/2$).
The Poisson bracket between the holonomy and flux is given by
\be
\begin{split}
\{h_e,h_{e'}\}_\gamma&=0,\\
\{P^i_e,P^j_{e'}\}_\gamma&=-\delta_{ee'}\ii f^{ij}_k P^k_e,\\
\{P^i_e,h_{e'}\}_{\gamma}&=\delta_{ee'}\frac{\tau_i}{2}h_e.
\end{split}
\label{eq:poisson1}
\ee
The classical phase space associated with each edge $e$ is given by $G \times Lie(G)$, which can be identified with the cotangent bundle $T^*G$ \cite{TThiemann_2001QSD}.
Consequently, the classical phase space of the whole graph $\gamma$ is then given by the direct product of cotangent bundles $T^*G$ over all edges $e$ in $\gamma$:
\be
\mathcal{M}_\gamma=\otimes_{e\in E(\gamma)}T^*G,
\ee
with its symplectic structure $\Omega_\gamma$ determined by the Poisson brackets in \eqref{eq:poisson1}.

Following Lemma 3.1 in \cite{TThiemann_2001}, each $T^*G$ can be identified with the complexified group $G^{\mathbb{C}}$ via the polar decomposition:
\be
g_e:=e^{- \frac{p^i_e\tau_i}{2}}h_e\in G^{\mathbb{C}}\cong T^*G,
\label{eq:complexified}
\ee
where $p^i_e$ is the dimensionless version of $P^i_e$ defined as
\be
p^i_e=\frac{P^i_e}{a^{n^D}}.
\ee
Here, $a$ is a constant with the dimension of length, and $n^D$ is the dimension of $P^i_e$.
The value of $n^D$ is determined by dimensional analysis of the kinetic term in the classical gauge field action in a $D+1$-dimensional spacetime manifold:
\be
S_{\mathrm{kin}}=-\frac{1}{\kappa}\int_{\mathbb{R}\times\Sigma}dtd^DxE^a_i(x)A_a^i(x).
\ee
Since the dimensionless $h_e$ and dimensional $P^i_e/\kappa$ are canonical coordinates,
their Poisson bracket has units of action.
Thus, the dimension of $P^i_e/\kappa$ is the same as that of the action.
For instance, in Yang-Mills theory $n^D$ is $D-3$, while in general relativity $n^D$ is $D-1$.

With this dimensionless variable. the Poisson brackets \eqref{eq:poisson1} modify to 
\be
\begin{split}
\{h_e,h_{e'}\}_\gamma&=0,\\
\{P^i_e/\kappa,P^j_{e'}/\kappa\}_\gamma&=-\delta_{ee'}\ii f^{ij}_k P^k_e/\kappa,\\
\{P^i_e/\kappa,h_{e'}\}_{\gamma}&=\delta_{ee'}\frac{\tau_i}{2}h_e.
\end{split}
\label{eq:poisson2}
\ee
The complexification is then archived by using the heat kernel coherent state transformation. 
The complexifier $C_\gamma$ on the graph $\gamma$ is given by 
\be
C_\gamma:=\frac{1}{2 \kappa a^{n_D}} \sum_{e \in E(\gamma)} \delta^{i j} P_i^e P_j^e.
\ee
Then, Lemma 3.1 in  \cite{TThiemann_2001} is realized as
\be
\begin{aligned}
h_e^{\mathbb{C}} & :=g_e=\sum_{n=0}^{\infty} \frac{1}{n!}\left\{h_e, C\right\}_{(n)} \\
& =\left[\sum_{n=0}^{\infty} \frac{1}{n!}\left(-p_j^e \frac{\tau_j}{2}\right)^n\right] h_e \\
& =e^{- \tau_j p_j^e / 2} h_e=: H_e h_e.
\end{aligned}
\ee

Following \cite{HALL1994103,Hall1997,TThiemann_2001}, the coherent state on the edge $e$ labeled by $g_e\in G^{\mathbb{C}}$ is defined as
\be
\psi^t_{g_e}(h_e):=\sum_\pi d_\pi e^{-\frac{t}{2} \lambda_\pi} \chi_\pi\left(g h^{-1}\right),
\label{eq:coherent_state}
\ee
where the summation is done over all the irreducible representations $\pi$ of $G$, 
$d_\pi$ is the dimension of the representation $\pi$, 
$\lambda_\pi$ is the eigenvalue of the Laplacian on $G$ in the representation $\pi$, 
$\chi_\pi$ is the character function of $\pi$, 
and $t=\frac{\hbar \kappa}{a^{n_D}}$ is a dimensionless parameter controlling the semiclassical limit.
The coherent state for the entire graph $\gamma$ labeled by $\{g_e\}_{e\in E(\gamma)}$ is given by the tensor product over edges:
\be
\Psi^t_{\gamma,\{g_e\}}(\{h_e\}):=\prod_{e\in E(\gamma)}\psi^t_{g_e}(h_e).
\ee

As shown in \cite{TThiemann_20013,TThiemann_2001}, for $SU(2)$ and $U(1)$ groups, the coherent states constructed above exhibits following semiclassical properties in the limit $t\rightarrow 0$:

1. \textbf{Peakedness property}: 
The probability density function
\be
\label{eq:peakedness}
p^t_{g_e}(h_e):=\frac{|\psi^t_{g_e}(h_e)|^2}{\|\psi^t_{g}\|^2},
\ee
peaks sharply at $h_e=u$ as $t\rightarrow 0$ where $g_e=H_eu$ is the polar decomposition of $g_e$.

2. \textbf{Ehrenfest property}:
The expectation values of the basic operators $\hat{h}_e$ and $\hat{P}^i_e$ on the coherent states approximate their classical counterparts in the semiclassical limit.

3. \textbf{Overlap property}:
The overlap function
\be
\label{eq:overlap}
 i^t(g_e,g'_e) :=\frac{|\langle\psi^t_{g_e}|\psi^t_{g'_e}\rangle|^2}{\|\psi^t_{g_e}\|^2\|\psi^t_{g'_e}\|^2}
\ee
decays rapidly as $t\rightarrow 0$ when $g_e$ and $g'_e$ are far apart in the classical phase space $T^*G$.

In \cite{PhysRevD.102.024083}, the leading order of the overlap function in the limit where $t\rightarrow 0$ and $g_e$ is close to $g'_e$ is explicitly computed for $G=SU(2)$, which plays a crucial role in the semiclassical analysis of the coherent state path integral.
For higher-rank gauge group like $SU(3)$, the coherent states constructed above are expected to retain the semiclassical properties,
but the explicit proofs become more involved due to the complexity of the representation theory. In the following sections, we will specialize this general construction to the $SU(3)$ case and detail the semiclassical properties.

\section{$SU(3)$ Coherent States}
\subsection{Representation of $SU(3)$ group and construction of coherent states}
In order to apply \eqref{eq:coherent_state} to construct the $SU(3)$ coherent states, the representation theory of $SU(3)$ group is essential.
The rank of $SU(3)$ group is $2$; thus, the Young tableau representing its irreducible representations can at most have two rows.
An irreducible representation $\pi$ of $SU(3)$ can be labeled Dynkin label $(p,q)$, where $p$ is the number of boxes in the first row of the Young tableau, and $q$ is the number of boxes in the second row.
In order to compute the coherent states in \eqref{eq:coherent_state}, we need to know the dimension and the eigenvalue of the Laplacian on $SU(3)$ in the representation $(p,q)$, as well as character of each irreducible representation $(p,q)$.

The dimension $d_{(p,q)}$ of the irreducible representation $(p,q)$ yields to the expression
\be
d_{(p,q)} = \frac{num}{den}
\ee
The numerator $num$ is given by the following method:
\begin{itemize}
\item Start from the upper left box of the Young tableau, and assign it the number $3$.
\item Move to the right box, and assign it the number obtained by adding $1$ to the number of the left box.
\item Move to the lower box, and assign it the number obtained by adding $-1$ to the number of the upper box.
\item Continue this process until all boxes are assigned a number.
\item Finally, take the product of all the assigned numbers to obtain the numerator $num$.
\end{itemize}
The denominator $den$ is given by the following method:
\begin{itemize}
\item Write in each box the number of boxes locating to its right plus the number of boxes below it plus $1$.
\item Take the product of all the assigned numbers to obtain the denominator $den$.
\end{itemize}
Therefore, the dimension $d_{(p,q)}$ is given by
\be
d_{(p,q)} = \frac{(p+1)(q+1)(p+q+2)}{2}.
\label{eq:dimension}
\ee

The specific meaning of the Laplacian is the quadratic Casimir operator $C_2$ of $SU(3)$ group. 
The general expression of this operator is given by 
\be
C_2 = \langle m, m+2\delta \rangle,
\ee
where $m$ is the highest weight of the representation, $\delta$ is half the sum of all positive roots, and $\langle\cdot,\cdot\rangle$ is the inner product in the weight space.
For the irreducible representation $(p,q)$ of $SU(3)$, the highest weight is given by
\be
m = p\lambda_1 + q\lambda_2,
\ee
where $\lambda_1$ and $\lambda_2$ are the fundamental weights of $SU(3)$ group.
The half sum of all positive roots is given by
\be
\delta = \lambda_1 + \lambda_2.
\ee
The inner products between the fundamental weights are given by
\be
\langle \lambda_1, \lambda_1 \rangle = \langle \lambda_2, \lambda_2 \rangle = \frac{1}{3}, \quad \langle \lambda_1, \lambda_2 \rangle = \frac{1}{6}.
\ee
Thus, the eigenvalue $\lambda_{(p,q)}$ of the Laplacian on $SU(3)$ in the representation $(p,q)$ is given by \cite{Iachello2015}
\be
\lambda_{(p,q)} = \frac{1}{3}(p^2 + q^2 + pq + 3p + 3q).
\label{eq:laplacian}
\ee 

A general element $g\in SU(3)$ is given by
\be
g = e^{\ii \phi_i \tau_i},
\ee
where $\phi_i\in \mathbb{R}$ and $\tau_i$ are the generators of $su(3)$ Lie algebra.
Notice that the unitary matrix $g$ can be diagonalized, the character $\chi_{(p,q)}(g)$ only depends on the eigenvalues of $g$.
Let the eigenvalues of $g$ be $\{e^{\ii \theta_1}, e^{\ii \theta_2}, e^{\ii \theta_3}\}$, where $\theta_1+\theta_2+\theta_3=0$.
Denote $\varepsilon_i$ as $e^{\ii \theta_i}$, and define $h_i$ as the number of columns of length $i$ in each tableau.
Then, $p = h_1 +h_2$, $q = h_2$, and $h_3=0$ for the irreducible representation $\pi_{(p,q)}$.
The character $\chi_{(p,q)}(g)$ in the irreducible representation $(p,q)$ is given by the Weyl character formula:
\be
\chi(\hat{G}_a)
=\left|\begin{array}{lll}
\varepsilon_1^{h_{1}+2} & \varepsilon_1^{h_{2}+1} & 1 \\
\varepsilon_2^{h_{1}+2} & \varepsilon_2^{h_{2}+1} & 1 \\
\varepsilon_3^{h_{1}+2} & \varepsilon_3^{h_{2}+1} & 1
\end{array}\right| \cdot\left|\begin{array}{lll}
\varepsilon_1^2 & \varepsilon_1 & 1 \\
\varepsilon_2^2 & \varepsilon_2 & 1 \\
\varepsilon_3^2 & \varepsilon_3 & 1
\end{array}\right|^{-1}.
\ee
By computing the determinants, the character $\chi_{(p,q)}(g)$ can be simplified to
\be
\begin{aligned}
\chi^{p, q}(\theta_1, \theta_2) & =-\frac{i}{s(\theta_1, \theta_2)}\left[-e^{i((p+1) \theta_2-(q+1) \theta_1)}\right. \\
& \left.+e^{i((p+1) \theta_1-(q+1) \theta_2)}\right.\\
&\left.+e^{-i(p+1)(\theta_1+\theta_2)}\left(e^{-i(q+1) \theta_1}-e^{-i(q+1) \theta_2}\right)\right. \\
&\left.+e^{i(q+1)(\theta_1+\theta_2)}\left(e^{i(p+1) \theta_2}-e^{i(p+1) \theta_1}\right)\right] \\
s(\theta_1, \theta_2) & =8 \sin \left(\frac{\theta_1-\theta_2}{2}\right) \sin \left(\frac{1}{2}(\theta_1+2 \theta_2)\right) \\
&\times \sin \left(\frac{1}{2}(2 \theta_1+\theta_2)\right).
\end{aligned}
\ee
Denote $J_1 = p+1$, $J_2 = q+1$, and then the character $\chi_{(p,q)}(g)$ can be rewritten as \cite{Baaquie_1988}
\be
\begin{aligned}
\chi^{J_1, J_2 }(\theta_1, \theta_2) & =-\frac{i}{s(\theta_1, \theta_2)}\left[-e^{i(J_1 \theta_2-J_2 \theta_1)}\right. \\
& \left.+e^{i(J_1 \theta_1-J_2 \theta_2)}\right.\\
&n\left.+e^{-iJ_1(\theta_1+\theta_2)}\left(e^{-iJ_2 \theta_1}-e^{-iJ_2 \theta_2}\right)\right.\\
&\left.+e^{iJ_2(\theta_1+\theta_2)}\left(e^{iJ_1 \theta_2}-e^{iJ_1 \theta_1}\right)\right] \\
s(\theta_1, \theta_2) & =8 \sin \left(\frac{\theta_1-\theta_2}{2}\right) \sin \left(\frac{1}{2}(\theta_1+2 \theta_2)\right) \\
&\times\sin \left(\frac{1}{2}(2 \theta_1+\theta_2)\right).
\end{aligned}
\label{eq:character}
\ee

With the dimension \eqref{eq:dimension}, the eigenvalue of the Laplacian \eqref{eq:laplacian}, and the character \eqref{eq:character} of each irreducible representation $(p,q)$ of $SU(3)$, we can construct the $SU(3)$ coherent state on the edge $e$ labeled by $g\in SU(3)^{\mathbb{C}}$ as
\be
\begin{split}
\psi_g^t(h)
&=\sum_\pi 
    d_\pi 
    e^{-\frac{t}{2} \lambda_\pi} \chi_\pi\left(g h^{-1}\right)
\\
&=\sum_{J_1=1}^{+\infty} \sum_{J_2=1}^{+\infty} 
    -
    \frac{J_1J_2(J_1+J_2)i}{2s(\theta_1, \theta_2)} \\
    &\times
    \exp{\left(
        -\frac{t}{6} \left(
            J_1^2+J_2^2+J_1J_2-3
        \right)
    \right)} 
    \\
    &\times\left[
        -e^{i(J_1 \theta_2-J_2 \theta_1)}
        +e^{i(J_1 \theta_1-J_2 \theta_2)}
    \right.\\
    &\left.
        +e^{-iJ_1(\theta_1+\theta_2)}\left(
            e^{-iJ_2 \theta_1}-e^{-iJ_2 \theta_2}
        \right)
    \right. \\
    &\left.
        +e^{iJ_2(\theta_1+\theta_2)}\left(
            e^{iJ_1 \theta_2}-e^{iJ_1 \theta_1}
        \right)
    \right], 
\end{split}
\label{eq:SU3_coherent_state}
\ee
where $\{e^{\ii \theta_1}, e^{\ii \theta_2}\}$ are any two of the eigenvalues of $gh^{-1}$, and the third eigenvalue is given by $e^{-\ii(\theta_1+\theta_2)}$ due to the unimodularity of $SU(3)$ group.
Since $g \in SU(3)^{\mathbb{C}}$, the eigenvalues $\{e^{\ii \theta_1}, e^{\ii \theta_2}\}$ are generally complex numbers.

\subsection{Asymptotic analysis of $SU(3)$ coherent states}
To analysis the behavior of this $SU(3)$ coherent state in the limit $t\rightarrow 0$, the Poisson summation formula is needed to convert the summation over discrete variables $J_1$ and $J_2$ into integrals over continuous variables.
One difficulty arises from the fact that $J_1$ and $J_2$ are both positive integers starting from $1$, rather than all integers.
The resolution is to rewrite the $J_1$ and $J_2$ to be their absolute values and extend the summation:
\be
\begin{split}
    \psi_g^t(h)
    &=\sum_{J_1=1}^{+\infty} \sum_{J_2=1}^{+\infty} 
    -
    \frac{|J_1||J_2|(|J_1|+|J_2|)i}{2s(\theta_1, \theta_2)} \\
    &\times
    \exp{\left(
        -\frac{t}{6} \left(
            |J_1|^2+|J_2|^2+|J_1||J_2|-3
        \right)
    \right)} 
    \\
    &\times\left[
        -e^{i(|J_1| \theta_2-|J_2| \theta_1)}
        +e^{i(|J_1| \theta_1-|J_2| \theta_2)}
    \right.\\
    &\left.
        +e^{-i|J_1|(\theta_1+\theta_2)}\left(
            e^{-i|J_2| \theta_1}-e^{-i|J_2| \theta_2}
        \right)
    \right. \\
    &\left.
        +e^{i|J_2|(\theta_1+\theta_2)}\left(
            e^{i|J_1| \theta_2}-e^{i|J_1| \theta_1}
        \right)
    \right]. 
\\
\end{split}
\ee
When $J_1=J_2=0$ the summand is zero; thus, the four times the summation can be written as the extension of the summation to all integers:
\be
\begin{split}
    \psi_g^t(h)
    &=\frac{1}{4}\sum_{J_1=-\infty}^{+\infty} \sum_{J_2=-\infty}^{+\infty}
    -
    \frac{|J_1||J_2|(|J_1|+|J_2|)i}{2s(\theta_1, \theta_2)} \\
    &\times
    \exp{\left(
        -\frac{t}{6} \left(
            |J_1|^2+|J_2|^2+|J_1||J_2|-3
        \right)
    \right)} 
    \\
    &\times\left[
        -e^{i(|J_1| \theta_2-|J_2| \theta_1)}
        +e^{i(|J_1| \theta_1-|J_2| \theta_2)}
    \right.\\
    &\left.
        +e^{-i|J_1|(\theta_1+\theta_2)}\left(
            e^{-i|J_2| \theta_1}-e^{-i|J_2| \theta_2}
        \right)
    \right. \\
    &\left.
        +e^{i|J_2|(\theta_1+\theta_2)}\left(
            e^{i|J_1| \theta_2}-e^{i|J_1| \theta_1}
        \right)
    \right] .
\\
\end{split}
\ee
Since the semiclassical limit $t\rightarrow 0$ also corresponds to the limit where $J_1$ and $J_2$ approach infinity, it is reasonable to rewrite the $J_i$ as $j_i/t$ where $1/t$ is defined as a uniform scale of the $J_i$.
Thus, the coherent state can be rewritten as
\be
\begin{split}
    \psi_g^t(h)
    &=-\frac{ie^{\frac{t}{2}}}{8t^3s(\theta_1, \theta_2)}\\
    &\times
    \sum_{j_1=-\infty}^{+\infty} \sum_{j_2=-\infty}^{+\infty}
    \left(|j_1||j_2|(|j_1|+|j_2|)\right) \\
    &\times
    \exp{\left(
        -\frac{1}{6t} \left(
            |j_1|^2+|j_2|^2+|j_1||j_2|
        \right)
    \right)} 
    \\
    &\times\left[
        -e^{\frac{i}{t}(|j_1| \theta_2-|j_2| \theta_1)}
        +e^{\frac{i}{t}(|j_1| \theta_1-|j_2| \theta_2)}\right.\\
        &\left.+e^{-\frac{i}{t}(|j_1|(\theta_1+\theta_2)+|j_2|\theta_1)}
        -e^{-\frac{i}{t}(j_1|(\theta_1+\theta_2)+|j_2|\theta_2)}\right.\\
        &\left.+e^{\frac{i}{t}(|j_2|(\theta_1+\theta_2)+|j_1|\theta_2)}
        -e^{\frac{i}{t}(|j_2|(\theta_1+\theta_2)+|j_1|\theta_1)}
    \right] .
\\
\end{split}
\label{eq:coherent_detail}
\ee
For convenience, the $\psi_g^t(h)$ can be decomposed into six terms, and the general form of each term is given by
\be
\begin{split}
    Term_i 
    &=-\frac{ie^{\frac{t}{2}}}{8t^3s(\theta_1, \theta_2)}\\
    &\times
    \sum_{j_1=-\infty}^{+\infty} \sum_{j_2=-\infty}^{+\infty}
    \left(|j_1||j_2|(|j_1|+|j_2|)\right) \\
    &\times
    \exp{\left(
        -\frac{1}{6t} \left(
            |j_1|^2+|j_2|^2+|j_1||j_2|
        \right)
    \right)}
    \mathcal{F}_i,
\end{split}
\ee
where $\mathcal{F}_i, i = 1,2,\cdots 6$ represents the six terms in the last factor in \eqref{eq:coherent_detail}.
The Poisson summation formula can be applied to each term separately. Takeing $Term_1$ as an example, we have  
\be
\begin{split}
    &\quad Term_1
    =-\frac{ie^{\frac{t}{2}}}{8t^3s(\theta_1, \theta_2)}\\
    &\times
    \sum_{j_1=-\infty}^{+\infty} \sum_{j_2=-\infty}^{+\infty}
    \left(|j_1||j_2|(|j_1|+|j_2|)\right) \\
    &\times
    \exp{\left(
        -\frac{1}{6t} \left(
            |j_1|^2+|j_2|^2+|j_1||j_2|
        \right)
    \right)}\\
    &\times
    \left(
        e^{\frac{i}{t}(|j_1| \theta_2-|j_2| \theta_1)}
    \right)
    \\
    &=-\frac{ie^{\frac{t}{2}}}{8t^3s(\theta_1, \theta_2)}\sum_{K_1=-\infty}^{+\infty} \sum_{K_2=-\infty}^{+\infty}\\
    &\times
    \int_{-\infty}^{+\infty} \int_{-\infty}^{+\infty} d j_1  d j_2
    \left(|j_1||j_2|(|j_1|+|j_2|)\right) \\
    &\times
    \exp{\frac{1}{t}\left(
        -\frac{\left(
            |j_1|^2+|j_2|^2+|j_1||j_2|
        \right)}{6} 
        +
        i(|j_1| \theta_2-|j_2| \theta_1)
    \right)}\\
    &\times
    \exp{
    \left(
        -i2\pi K_1 \frac{j_1}{t}
    \right)
    }
    \exp{
    \left(
        -i2\pi K_2 \frac{j_2}{t}
    \right)
    }.
\end{split}
\ee
The absolute values in the integrand can be removed by dividing the integration domain into four quadrants:
\be
\begin{split}
    &\quad Term_1
    =-\frac{ie^{\frac{t}{2}}}{8t^3s(\theta_1, \theta_2)}\sum_{K_1=-\infty}^{+\infty} \sum_{K_2=-\infty}^{+\infty}\\
    &\times
    \int_{0}^{+\infty} \int_{0}^{+\infty} d j_1  d j_2
    \left(j_1j_2(j_1+j_2)\right) \\
    &\times
    \exp{\frac{1}{t}\left(
        -\frac{\left(
            j_1^2+j_2^2+j_1j_2
        \right)}{6} 
        +
        i(j_1 \theta_2-j_2 \theta_1)
    \right)}\\
    &\times
    \left[
    \exp{
    \left(
        -i2\pi K_1 \frac{j_1}{t}
    \right)
    }
    +
    \exp{
    \left(
        +i2\pi K_1 \frac{j_1}{t}
    \right)
    }
    \right]\\
    &\times
    \left[
    \exp{
    \left(
        -i2\pi K_2 \frac{j_2}{t}
    \right)
    }
    +
    \exp{
    \left(
        +i2\pi K_2 \frac{j_2}{t}
    \right)
    }
    \right]
    .
\end{split}
\ee
Considering the summation over $K_1$ and $K_2$ are done from $-\infty$ to $\infty$, the $K_i$ in the above expression can be freely replaced by $-K_i$.
Thus, the $Term_1$ can be simplified to
\be
\begin{split}
    &\quad Term_1
    =-\frac{ie^{\frac{t}{2}}}{8t^3s(\theta_1, \theta_2)}\sum_{K_1=-\infty}^{+\infty} \sum_{K_2=-\infty}^{+\infty}\\
    &\times
    \int_{0}^{+\infty} \int_{0}^{+\infty} d j_1  d j_2
    \left(j_1j_2(j_1+j_2)\right) \\
    &\times
    \exp{\frac{1}{t}\left(
        -\frac{\left(
            j_1^2+j_2^2+j_1j_2
        \right)}{6} 
        +
        i(j_1 \theta_2-j_2 \theta_1)
    \right)}\\
    &\times
    \exp{
    \left(
        i2\pi K_1 \frac{j_1}{t}
    \right)
    }
    \exp{
    \left(
        i2\pi K_2 \frac{j_2}{t}
    \right)
    }.
    %&=-\frac{ie^{\frac{t}{2}}}{8t^3s(\theta_1, \theta_2)}\sum_{K_1=-\infty}^{+\infty} \sum_{K_2=-\infty}^{+\infty}\\
    %&\times
    %\int_{0}^{+\infty} \int_{0}^{+\infty} d j_1  d j_2
    %\left(j_1j_2(j_1+j_2)\right) \\
    %&\times
    %\exp{\frac{1}{t}\left(
    %    -\frac{\left(
    %        j_1^2+j_2^2+j_1j_2
    %    \right)}{6}
    %    \right)}\\ 
    %&\times\exp{\frac{1}{t}\left(
    %    i(j_1 (\theta_2+2\pi K_1)-j_2 (\theta_1+2\pi K_2 ))
    %\right)}.
\end{split}
\ee
The rest of the five terms can be treated in the same way. The general form of these six terms after applying the Poisson summation formula is given by
\be
\begin{split}
    &\quad Term_i
    =\pm\frac{ie^{\frac{t}{2}}}{8t^3s(\theta_1, \theta_2)}\sum_{K_1=-\infty}^{+\infty} \sum_{K_2=-\infty}^{+\infty}\\
    &\times
    \int_{0}^{+\infty} \int_{0}^{+\infty} d j_1  d j_2
    \left(j_1j_2(j_1+j_2)\right) \\
    &\times
    \exp{\frac{1}{t}\left(
        -\frac{\left(
            j_1^2+j_2^2+j_1j_2
        \right)}{6} 
        +
        i(j_1 A^i_1-j_2 A^i_2)
    \right)},\\
\end{split}
\ee
where
\be
\label{eq:A_terms}
\begin{aligned}
A^1_1 & = \theta_2 + 2\pi K_1, & A^1_2 & = -\theta_1 + 2\pi K_2, \\
A^2_1 & = \theta_1 + 2\pi K_1, & A^2_2 & = -\theta_2 + 2\pi K_2, \\
A^3_1 & = -(\theta_1 + \theta_2) + 2\pi K_1, & A^3_2 & = \ -\theta_1 + 2\pi K_2, \\
A^4_1 & = \ -(\theta_1 + \theta_2)+ 2\pi K_1, & A^4_2 & = - \theta_2 + 2\pi K_2, \\
A^5_1 & = \theta_2 + 2\pi K_1, & A^6_2 & = \theta_1 + \theta_2 + 2\pi K_2, \\
A^6_1 & = \theta_1 + 2\pi K_1, & A^5_2 & = \theta_1 + \theta_2 + 2\pi K_2, \\
\end{aligned}
\ee
and the plus or minus sign in front of each term corresponds to the six terms in \eqref{eq:coherent_detail} respectively.
For each term, the integrals over $j_1$ and $j_2$ are both Gaussian integrals and can be explicitly computed.
\be
\begin{split}
Term_g 
=& 
\int^{\infty}_{0} \int^{\infty}_{0}dj_1 dj_2
(j_1 j_2(j_1+j_2))\\
\times&
\exp{
    \frac{1}{t}
    \left[
        -\frac{1}{6} \left(
            j_1^2+j_2^2+j_1j_2
        \right)
        +
        i(j_1 A_1+j_2 A_2)
    \right]
},\\
=&
it^3
(\frac{\partial^2}{\partial A_1^2} \frac{\partial}{\partial A_2}+\frac{\partial}{\partial A_1} \frac{\partial^2}{\partial A_2^2})
\int^{\infty}_{0} \int^{\infty}_{0}dj_1 dj_2\\
\times&
\exp{
    \frac{1}{t}
    \left[
        -\frac{1}{6} \left(
            j_1^2+j_2^2+j_1j_2
        \right)
        +
        i(j_1 A_1+j_2 A_2)
    \right]
}.
\end{split}
\ee
The integrals over $j_1$ and $j_2$ are coupled Gaussian integrals, which can be computed by diagonalizing the quadratic form in the exponent.
Let $q_1 = \frac{1}{\sqrt{2}}\left( j_1-j_2 \right)$, and $q_2 = \frac{1}{\sqrt{2}}\left( j_1+j_2 \right)$, and then the integral can be rewritten as
\be
\begin{split}
Term_g
=&
it^3
(\frac{\partial^2}{\partial A_1^2} \frac{\partial}{\partial A_2}+\frac{\partial}{\partial A_1} \frac{\partial^2}{\partial A_2^2})\\
\times&
\int^{\infty}_{0} dq_1
\exp{
    \frac{1}{t}
    \left(    
         -\frac{q_1^2}{12}
         -\frac{\ii q_1(A_2-A1)}{\sqrt{2}}
    \right)
}\\
\times&
\int^{\infty}_{0} dq_2
\exp{
    \frac{1}{t}
    \left(    
         -\frac{q_2^2}{4}
         +\frac{iq_2(A_2+A1)}{\sqrt{2}}
    \right)
}
\end{split}
\ee
Then, the problem reduces to compute the integrals of the form
\be
I_1(a,b)
=
\int^{\infty}_{0} dq_1
\exp{
    \frac{1}{t}
    \left(    
         -\frac{q_1^2}{12}
         +\frac{\ii q_1\mathcal{A}}{\sqrt{2}}
    \right)
}.
\ee
The saddle-point of the logarithm of the integrand is given by $q_1 =  3\sqrt{2}\ii\mathcal{A}$.
The integration contour can be deformed to pass through the saddle-point along the steepest descent direction (Lefschetz thimble); thus, the integral yields
\be
\label{eq:I_ab}
\begin{split}
I_1(a,b)&=
\int_{0}^{\ii 3\sqrt{2}(\mathcal{A})} 
\exp{
    \frac{1}{t}
    \left(    
         -\frac{q_1^2}{12}
         +\frac{\ii q_1\mathcal{A}}{\sqrt{2}}
    \right)
}
dq_1\\
&+\int_{thimble} 
\exp{
    \frac{1}{t}
    \left(    
         -\frac{q_1^2}{12}
         +\frac{\ii q_1\mathcal{A}}{\sqrt{2}}
    \right)
}
dq_1,
\end{split}
\ee
where the $thimble$ is the line starting from the saddle-point and parallel to the real axis.
Denoting $q_1$ as $r+3\sqrt{2}\ii\mathcal{A}$, the second integral in \eqref{eq:I_ab} becomes
\be
\begin{split}
    &\int_{thimble} 
    \exp{
        \frac{1}{t}
        \left(    
             -\frac{q_1^2}{12}
             +\frac{\ii q_1\mathcal{A}}{\sqrt{2}}
        \right)
    }
    dq_1
    \\
    =&
    \int_{0}^\infty 
    \exp{
        \frac{1}{t}
        \left(    
             -\frac{r^2}{12}
             -\frac{3}{2}
            \mathcal{A}^2
        \right)
    }
    dr
    \\
    =&
    \sqrt{
        3\pi
    }
    \sqrt{t}
    \exp\left(
        -\frac{3(\mathcal{A})^2}{2t}    
    \right).
\end{split}
\ee
The first integral in \eqref{eq:I_ab} can be computed directly, yielding
\be
\begin{split}
    &\int_{0}^{i 3\sqrt{2}(\mathcal{A})} 
    \exp{
        \frac{1}{t}
        \left(    
             -\frac{q_1^2}{12}
             +\frac{\ii q_1\mathcal{A}}{\sqrt{2}}
        \right)
    }
    dq_1
    \\
    =&
    -\ii \sqrt{3\pi t}
    \exp\left(
        -\frac{3\mathcal{A}^2}{2t}    
    \right)
    \\
    \times&
    \operatorname{Erfi}\left(
        \frac{
            -
            \sqrt{\frac{3}{2}}
            \mathcal{A}
        }{
            \sqrt{t}
        }
    \right).
\end{split}
\ee
Thus, the integral $I(a,b)$ yields
\be
\begin{split}
I_1(a,b)&=
\sqrt{3\pi t}
\exp\left(
    -\frac{3\mathcal{A}^2}{2t}    
\right)\\
&\times
\left(
    1
    +\ii
    \operatorname{Erfi}\left(
        \sqrt{\frac{3}{2t}}
        \mathcal{A}
    \right)
\right).
\end{split}
\ee
The error function $\operatorname{Erfi}(x)$ has the asymptotic behavior
\be
\operatorname{Erfi}(x)
\sim
\frac{e^{x^2}}{\sqrt{\pi} x}
\left(
    1
    +
    \sum_{m=1}^{\infty} \frac{(2 m-1)!!}{\left(2 x^2\right)^m}
\right)
\ee
for $|x|\rightarrow \infty$.
Thus, in the limit $t\rightarrow 0$, the integral parts in $Term_g$ yield
\be
\begin{split}
    I_g=&\int^{\infty}_{0} 
    \exp{
        \frac{1}{t}
        \left(    
             -\frac{q_1^2}{12}
             +\frac{iq_1(A_2-A1)}{\sqrt{2}}
        \right)
    }
    dq_1
    \\
    \times
    &
    \int^{\infty}_{0}
    \exp{
        \frac{1}{t}
        \left(    
             -\frac{q_2^2}{4}
             +\frac{iq_2(A_2+A_1)}{\sqrt{2}}
        \right)
    }
    dq_2
    \\
    \sim&
\sqrt{
    3
}
\pi
t
e^{
    -\frac{2\left(A_1^2-A_1A_2+A_2^2\right)}{t}    
}
\\
+&
i
\frac{
    \sqrt{6\pi} t^{3/2}
}{
    A_1+A_2
}
e^{
    -\frac{3(A_1-A_2)^2}{2t}    
}\\
\times& 
\left(
    1
    +
    \sum_{m=1}^{\infty} 
        \frac{
            t^m
            (2 m-1)!!
        }{
            \left(
                    A_2+A_1
            \right)^{2m}
        }
\right)
\\
+&
i
\frac{\sqrt{2\pi}t^{3/2}}{A_1-A_2}
e^{-\frac{(A_1+A_2)^2}{2t}}\\
\times&
\left(
    1
    +
    \sum_{m=1}^{\infty} 
        \frac{
            t^m
            (2 m-1)!!
        }{
            3^m
            \left(
                    A_2-A_1
            \right)^{2m}
        }
\right)
\\
-&
\frac{2t^2}{A_1^2-A_2^2}
\left(
    1
    +
    \sum_{m=1}^{\infty} 
        \frac{
            t^m
            (2 m-1)!!
        }{
            \left(
                    A_2+A_1
            \right)^{2m}
        }
\right)\\
\times&
\left(
    1
    +
    \sum_{m=1}^{\infty} 
        \frac{
            t^m
            (2 m-1)!!
        }{
            3^m
            \left(
                    A_2-A_1
            \right)^{2m}
        }
\right).
\end{split}
\ee
The first three terms in the above expression are all exponentially suppressed in the limit $t\rightarrow 0$; thus, the leading contribution to $I_g$ comes from the last term in the above expression.
However, in the following discussion we will show that the contribution of this term to the coherent state vanishes after summing over all the six terms.
Denote the last term in the above expression as $L^{(0)}(A_1,A_2)$, and then the contribution of this part to $Term_g$ is given by
\be
\begin{split}
    Term_g
    \sim
    it^3
    (\frac{\partial^2}{\partial A_1^2} \frac{\partial}{\partial A_2}+\frac  {\partial}{\partial A_1} \frac{\partial^2}{\partial A_2^2})
    L^{(0)}(A_1,A_2)
\end{split}
\ee
Applying \eqref{eq:A_terms} $Term_g$, we can compute the leading contributions of all six terns of the coherent state.
Denote these terms as $Term_i^{(0)}$, and it can be shown that if we convert $K_1$ to be $-K_2$ and $K_2$ to be $-K_1$ in $Term_2^{(0)}$, $Term_5^{(0)}$, and $Term_6^{(0)}$, then the summation of all six terms vanishes.
Therefore, the leading contribution to the coherent state comes from the next-to-leading-order terms in $I_g$.
Among the first three exponentially suppressed terms in $I_g$, the first term contains the lowest order of $t$; thus, this term provides the leading contribution in \eqref{eq:coherent_detail}.

The summation over all six terms is given by \eqref{eq:summand_final}. The summand in \eqref{eq:summand_final} has a common factor $e^{(-8\pi^2(K_1^2-K_1K_2+K_2^2)/t)}$.
In the limit $t\rightarrow 0$, the leading contribution to the coherent state corresponds to the case when $K_1^2-K_1K_2+K_2^2$ takes the smallest value (according to AM-GM inequality ), which is $K_1=0$ and $K_2=0$.
Therefore, the leading contribution to the coherent state in the limit $t\rightarrow 0$ is given by
\be
\begin{split}
\psi_g^t(h)&
\sim
\frac{ 24 e^{\frac{t}{2}}\sqrt{3}\pi}{t^2}
e^{
    -
    \frac{
        2
        \left(
            \theta_1^2
            +
            \theta_1
            \theta_2
            +
            \theta_2^2
        \right)
    }{
        t
    }
}\\
\times&
\frac{
    (\theta_1-\theta_2)
    (2\theta_1+\theta_2)
    (\theta_1+2\theta_2)
}{
    s(\theta_1, \theta_2)
}
\\
=&
\frac{ 24 \sqrt{3}\pi e^{\frac{t}{2}} }{ t^2 }
e^{
        -
        \frac{
            2
            \left(
                \theta_1^2
                +
                \theta_1
                \theta_2
                +
                \theta_2^2
            \right)
        }{
            t
        }
    }\\
\times&
\frac{
    1
}{
    \operatorname{sinc} \left(\frac{\theta_1-\theta_2}{2}\right) 
    \operatorname{sinc} \left(\frac{\theta_1+2 \theta_2}{2}\right) 
    \operatorname{sinc} \left(\frac{2 \theta_1+\theta_2}{2}\right)
}.
\end{split}
\label{eq:coherent_final_asymptotics}
\ee

\subsection{Solve matrix exponential for $SU(3)$}
The expression \eqref{eq:coherent_final_asymptotics} is given in terms of the eigenvalues $\theta_1$ and $\theta_2$ of the group element $h\in SU(3)$.
To express the coherent state in terms of the matrix elements of $h$, we need to solve the eigenvalues $\theta_1$ and $\theta_2$ from the matrix $h$
The general form of a complexified $SU(3)$ group element in its fundamental representation is given by 
\be
h = \exp(-\ii v^a \lambda_a), 
\ee
where $\lambda_a, a=1,2,\cdots,8$ are the Gell-Mann matrices normalized as $\operatorname{tr}(\lambda_a \lambda_b) = 2 \delta_{ab}$, and $v^a\in \mathbb{C}$. The commutation relation of the Gell-Mann matrices is given by
\be
[\lambda_a, \lambda_b] = 2\ii f_{abc} \lambda_c,
\ee
where $f_{abc}$ are the structure constants of the $SU(3)$ Lie algebra.
Denote $\Sigma$ as the normalized generator $v^a\lambda_a/\langle v \rangle$, where $\langle v \rangle^2 = \frac{1}{2} \operatorname{tr}(v^a\lambda_a v^b\lambda_b) $.
To be notice is that the $\langle v \rangle$ does not have to be real for complexified $SU(3)$ group elements.
Therefore, we do not use the notation $|v|$ to avoid confusion.
The relation between previously mentioned $\theta_1, \quad\theta_2$ and $\Sigma$ can be given by Cayley-Hamilton relation.
The characteristic equation of the matrix $\Sigma$ is given by
\be
P(\lambda) = \det(\lambda I - \Sigma) = \lambda^3 - \frac{1}{2}\operatorname{tr}(\Sigma^2)\lambda - \operatorname{det}(\Sigma) = 0.
\ee
Considering the normalization of $\Sigma$, we have $\operatorname{tr}(\Sigma^2) = 2$.
According to Cayley-Hamilton relation, the matrix $\Sigma$ itself satisfies the characteristic equation, i.e.,
\be
\Sigma^3 - \Sigma - \operatorname{det}(\Sigma) I = 0.
\label{eq:cayley_hamilton}
\ee
The $\Sigma$ can diagonalized as $\Sigma = U \operatorname{diag}(z_1, z_2, z_3) U^{-1}$, where $\sum_i z_i =0$.
Then, Cayley-Hamilton relation indicates that the eigenvalues $z_i$ satisfy the equation
\be
z_i^3 - z_i - \operatorname{det}(\Sigma) = 0.
\ee
The three roots of the above equation can be expressed as \cite{Kaiser2022}
\be
\begin{split}
z_1 &= \frac{2}{\sqrt{3}} \cos\left(\phi\right), \\
z_2 &= -\sin\left(\phi\right) - \frac{\cos\left(\phi\right)}{\sqrt{3}}, \\
z_3 &= \sin\left(\phi\right) - \frac{\cos\left(\phi\right)}{\sqrt{3}},
\end{split}
\ee
where $\psi=\frac{1}{3} \arccos \frac{3 \sqrt{3} \eta}{2}$, and $\eta = \operatorname{det}(\Sigma)$.
The eigenvalues $\theta_1$ and $\theta_2$ of the group element $h$ are $- \ii \langle v \rangle z_i$ where $i=1,2$.
Thus, the coherent state \eqref{eq:coherent_final_asymptotics} can be expressed in terms of the matrix elements of $gh^{-1}$ by
\be\label{eq:coherent_semi_final_1}
\psi^t_g(h)
\sim
\frac{ 24 \sqrt{3}\pi e^{\frac{t}{2}} }{ t^2 }
e^{
        \frac{
            2
            \langle v \rangle^2
        }{
            t
        }
}
\Phi(\langle v \rangle, \eta),
\ee
with
\[
\begin{split}
    \Phi(\langle v \rangle, \eta)
    =&
    \frac{
    \left(\langle v \rangle\sin{(\eta)}\right) 
}{
    \operatorname{sinh} \left(\langle v \rangle\sin{(\eta)}\right)  
}
\frac{
    \left(\langle v \rangle\frac{\sqrt{3}\cos{(\eta)}-\sin{(\eta)}}{2}\right)
}{
    \operatorname{sinh} \left(\langle v \rangle\frac{-\sqrt{3}\cos{(\eta)}+\sin{(\eta)}}{2}\right) 
}\\
\times&\frac{
    \left(\langle v \rangle\frac{\sqrt{3}\cos{(\eta)}+\sin{(\eta)}}{2}\right)
}{
    \operatorname{sinh} \left(\langle v \rangle\frac{\sqrt{3}\cos{(\eta)}+\sin{(\eta)}}{2}\right)
}.
\end{split}
\]
Thus, we have expressed the semiclassical limit of the coherent state in terms of $\langle v \rangle$ and $\eta$, which can be computed directly from the matrix elements of $gh^{-1}$.
For convenience, we call the norm of the generator as the \textit{magnitude} parameter, and the determinant of the normalized generator as the \textit{determinant} parameter.

\section{Semiclassical peakedness property and numerical verification}

In the semiclassical limit $t \rightarrow 0$, the $SU(3)$ coherent state $\psi^t_g(h)$ should satisfy peakedness property \eqref{eq:peakedness}. 
This section provides a numerical analysis to verify this property of the coherent state in its semiclassical limit derived in the previous section.

Let $g = H u$ be the polar decomposition of $g$, where the $H$ is Hermitian and $u$ is unitary. 
Therefore, the coherent state becomes 
\be
\psi^t_g(h) = \psi^t_{H}(h u^{-1}).
\ee
The denominator in \eqref{eq:peakedness} can be computed by using orthogonality relation:
\be
\label{eq:orthogonality}
\int_G d \mu_H(h) \overline{\pi(h)_{m n}} \pi^{\prime}(h)_{m^{\prime} n^{\prime}}=\frac{1}{d_\pi} \delta_{\pi \pi^{\prime}} \delta_{m m^{\prime}} \delta_{n n^{\prime}},
\ee
which yields
\be\label{eq:norm_coherent}
\begin{split}
\|\psi^t_g\|^2
= &
    \sum_\pi  d_\pi e^{-t \lambda_\pi} 
    \operatorname{Tr}(H^2)
\\
= &
\psi^{2t}_{H^2}(1),
\end{split}
\ee
where the detail is given in \eqref{eq:norm_calculation}.
The numerator in \eqref{eq:peakedness} can be computed by using the semiclassical limit of the coherent state derived in the previous section.
Assume $\langle v \rangle$ and $\eta$ as the magnitude and determinant parameters of $Huh^{-1}$, respectively, and $|v'|$ and $\eta'$ as the magnitude and determinant parameters of $H$, respectively, then \eqref{eq:peakedness} in semiclassical limit can be expressed as\footnote{
    The square of the magnitude parameter $\langle v'\rangle$ for Hermitian matrix $H$ is real, while the square of the magnitude parameter $\langle v \rangle$ for complexified $Huh^{-1}$ is complex in general. Since the nominator of the \eqref{eq:peakedness_semi} involves the square of the absolute value of the coherent state, only the real part of $\langle v \rangle^2$ contributes to the exponential factor in the nominator.
}
\be
\label{eq:peakedness_semi}
\begin{split}
p^t_{g}(h)=&\frac{|\psi^t_{g}(h)|^2}{\|\psi^t_{g}\|^2}\\
\sim&
\frac{96\sqrt{3}\pi}{t^2}
e^{
        \frac{
            4
            \left(\mathbf{Re}(\langle v \rangle)^2-|v'|^2\right)
        }{
            t
        }
}
\frac{|\Phi(\langle v \rangle, \eta)|^2}{\Phi(2|v'|, 8\eta')}.
\end{split}
\ee
In the limit $t\rightarrow 0$, the above expression is dominated by the exponential factor $e^{\frac{4(\mathbf{Re}(\langle v \rangle)^2-\langle v' \rangle^2)}{t}}$.
The function $p^t_g(h)$ reaches its maximum when $\mathbf{Re}(\langle v \rangle)^2 - \langle v' \rangle^2$ reaches its maximum.
However, to express the magnitude parameter $\langle v \rangle$ as a function of the parameters of $H$, $u$, and $h$ is complicated since the Baker-Campbell-Hausdorff formula is involved in the production of the matrix exponentials.

To numerically analyze the peakedness properties of the coherent state, we use Markov chain Monte Carlo (MCMC) method to probe the position of the peak. 
To compute the magnitude parameter of $Huh^{-1}$, we assume
\be
\begin{split}
H &= \exp{\left( v^a \lambda^a \right)},\\
u &= \exp{\left(  \ii w^a \lambda^a \right)},\\
h &= \exp{\left(  \ii (w^a+\delta w^a) \lambda^a \right)},\\
\end{split}
\ee 
where $v^a$, $w^a$, and $\delta w^{a}$ are real numbers.
Denote $W$ as $\ln{(Huh^{-1})}$, and $M$ as $\ln{(H)}$.
The magnitude parameters are given by
\be
\begin{split}
\langle v \rangle^2 = \operatorname{Tr}(W^2)/2, \\
\langle v' \rangle^2 = \operatorname{Tr}(M^2)/2. \\
\end{split}
\ee
Randomly choose $v^a$ and $w^a$, the Boltzmann factor $e^{\frac{4(\mathbf{Re}(\langle v \rangle)^2 - |v'|^2)}{t}}$ can be used to sample $\delta w^a$, where the factor $\beta = \frac{4}{t}$ plays the role of the inverse temperature in statistical mechanics. One would expect that in the \textit{low temperature} limit $t\rightarrow 0$, the sampled $\delta w^a$ would cluster around the point where $\mathbf{Re}(\langle v \rangle)^2 - |v'|^2$ reaches its maximum.
Using the Metropolis-Hastings algorithm, the sampled Markov chain will converge to the distribution $e^{\frac{4(\mathbf{Re}(\langle v \rangle)^2 - |v'|^2)}{t}}$.
Thus, the peakedness property can be numerically analyzed by analyzing the mean value and covariance property of the sampled Markov chain.
Fig\ref{fig:mcmc1} is the triangle plot showing the mean values and covariance among any pair of the eight $\delta w^a$ for a random choice of $v^a$ and $w^a$.
The result shows that at least locally the maximum value of $4(\mathbf{Re}(\langle v \rangle)^2 - |v'|^2)$ appears at $\delta w^a = 0$ for all $a=1,2,\cdots,8$, when $\frac{4}{t}$ reaches $1\times10^7$. 
That is to say, the coherent state $\psi^t_g(h)$ reaches its maximum at $h=u$ for a fixed $g=Hu$ at the limit $t\rightarrow 0$, which fulfills the expected peakedness property \eqref{eq:peakedness}. 
The code for the MCMC sampling and analysis is available at \cite{gitcode}.
One additional remark is that the function $4(\mathbf{Re}(\langle v \rangle)^2 - |v'|^2)$ may have multiple local maximum points since the $u$ and $h$ contains periodic structure, i.e., $\exp{(\ii \ \cdot\ )}$. However, to establish a one-to-one isomorphism between the parameter space and the $SU(3)$ group elements, it is necessary to restrict the parameters $\delta w^a$ to a fundamental domain within a single period of this cycle. Consequently, when the parameters are confined to this restricted range, only one of these local maxima is relevant, and the peakedness property of the coherent states remains well defined and unaffected.

\section{Peakedness of the overlap function in semiclassical limit and numerical verification}

The overlap function between two coherent states $\psi^t_g$ and $\psi^t_{g'}$ is defined by \eqref{eq:overlap}. There are two core tasks involved in this section: analyzing the peakedness property of the overlap function in the semiclassical limit $t\rightarrow 0$, and computing its explicit form in the combined limit where $t\rightarrow 0$ and $g\rightarrow g'$.
The latter is particularly crucial, as it underpins the effective dynamics from the coherent state path integral formulation of LQG \cite{PhysRevD.102.024083}. 

To express the overlap function in a more tractable form, using the polar decomposition $g = Hu$ and $g'=H'u'$, the denominator and nominator of \eqref{eq:overlap} can be simplified through \eqref{eq:norm_coherent} and \eqref{eq:orthogonality} respectively.
By \eqref{eq:norm_coherent}, $\|\psi^t_{g}\|$ and $\|\psi^t_{g'}\|$ in the denominator of \eqref{eq:overlap} yield
\be
\begin{split}
    \|\psi^t_{g}\|^2 \|\psi^t_{g'}\|^2 =
    \psi^{2t}_{H^2}(1) \psi^{2t}_{{H'}^2}(1).
\end{split}
\ee
Similarly, through orthogonality relation \eqref{eq:orthogonality}, the inner product in the nominator of \eqref{eq:overlap} is evaluated as
\be
\begin{split}
\langle\psi^t_{g}|\psi^t_{g'}\rangle &= \psi^{2t}_{HH'}(h)\\
&=\sum_\pi d_\pi e^{-t \lambda_\pi} \chi_\pi\left(HH' h^{-1}\right)
\end{split}
\ee
where $h = u u'^{-1}$. 
To analyze the limits of the overlap function, we parametrize the mentioned group elements as
\be
\begin{split}
H &= \exp{(v^a \lambda^a)},\\
H' &= \exp{((v^a+\delta v^a) \lambda^a)},\\
u &= \exp{( i w^a \lambda^a)},\\
u' &= \exp{( i (w^a+\delta w^a) \lambda^a)}.\\
\end{split}
\ee
Now, define the matrix $W=\ln{(HH'u'u^{-1})}$, and let $M$ be $\ln(H')$.
Let $\langle \tilde{v} \rangle$ and $\tilde{\eta}$ be the magnitude parameter and the determinant parameter of $HH'u'u^{-1}$ respectively, and let $\langle v' \rangle$ and $\eta'$ be those of $H'$.
The square of the magnitude parameters are given by $\langle \tilde{v} \rangle^2 = \operatorname{Tr}(W^2)/2$, and 
$\langle v' \rangle^2 = \operatorname{Tr}(M^2)/2$. 
In the limit $t\rightarrow 0$, the leading order expression of the overlap function becomes 
\be
\begin{split}
& i^{t}(g,g') =
\frac{|\psi^{2t}_{HH'}(h)|^2}{\psi^{2t}_{H^2}(1) \psi^{2t}_{{H'}^2}(1)}\\
\sim&
e^{\frac{
        Re(\langle \tilde{v} \rangle)^2-4\langle v \rangle^2+4\langle v' \rangle^2
    }{
        t
    }
}
\frac{
|\Phi(\langle \tilde{v} \rangle, \tilde{\eta})|^2
}{
\Phi(2\langle v' \rangle, 8\eta')
\Phi(2\langle v \rangle, 8\eta)
},
\end{split}
\ee
which is dominated by the exponential factor $e^{\frac{Re(\langle \tilde{v} \rangle)^2-4\langle v \rangle^2+4\langle v' \rangle^2}{t}}$.
Consequently, the overlap function reaches its maximum when $Re(\langle \tilde{v} \rangle)^2-4\langle v \rangle^2+4\langle v' \rangle^2$ is maximized.
However, similar to the previous section, expressing $\langle \tilde{v} \rangle$ as a function of the parameters of $H$, $H'$, $u$, and $u'$ is complicated due to the involvement of the Baker-Campbell-Hausdorff formula in the production of the matrix exponentials.
Thus, we again resort to numerical methods to analyze the peakedness property of the overlap function.

To numerically verify the peakedness property of the overlap function, we again use MCMC method to sample $\delta v^a$ and $\delta w^a$ according to the Boltzmann factor $e^{\beta(Re(\langle \tilde{v} \rangle)^2-4\langle v \rangle^2+4\langle v' \rangle^2)}$, where the factor $\beta = \frac{1}{t}$ plays the role of the inverse temperature in statistical mechanics. One would expect that in the \textit{Low temperature} limit $t\rightarrow 0$, the sampled $\delta v^a$ and $\delta w^a$ would cluster around the point where $Re(\langle \tilde{v} \rangle)^2-4\langle v \rangle^2+4\langle v' \rangle^2$ reaches its maximum.
Using the Metropolis-Hastings algorithm, the sampled Markov chain will converge to the prementioned distribution.
After a burn-in period of $10^5$ steps, we collected $20^6$ samples to ensure statistical reliability.
The results, visualized by triangle plots Fig.\ref{fig:mcmc_overlap}, show the mean values and covariance among any pair of the eight $\delta v^a$ and eight $\delta w^a$ for a random choice of the initial $v^a$ and $w^a$.
The result shows that at least locally the maximum value of $Re(\langle \tilde{v} \rangle)^2-4\langle v \rangle^2+4\langle v' \rangle^2$ appears at $\delta v^a = 0$ and $\delta w^a = 0$ for all $a=1,2,\cdots,8$, when $\frac{1}{t}$ reaches $10^7$. 
This result confirms that the overlap function $i^t(g,g')$ is sharply peaked at $g=g'$ in the semiclassical limit, fulfilling the desired peakedness property.
Similar to the previous case, the function $Re(\langle \tilde{v} \rangle)^2-4\langle v \rangle^2+4\langle v' \rangle^2$ may possess multiple local maxima due to the periodic structure of the group elements $u$ and $u'$.  However, by restricting the parameters $\delta w^a$ and $\delta v^a$ to a fundamental domain --- which is necessary to maintain a isomorphism between the parameter space and the  $SU(3)$ group elements --- we ensure that these additional local maxima do not affect the global peakedness property. Thus, the coherent states remain sharply peaked in the intended sense.
The code for the MCMC sampling and analysis is available at \cite{gitcode}.

\section{Overlap amplitude in the Combined Limit}
The overlap amplitude $k^t(g,g')$ is defined by
\be
k^t(g,g') :=\frac{\langle\psi^t_{g}|\psi^t_{g'}\rangle^2}{\|\psi^t_{g}\|^2\|\psi^t_{g'}\|^2},
\ee
and the combined limit is given by first taking the limit $t\rightarrow 0$ and then taking $g\rightarrow g'$.
In order to compute the leading order expression of the overlap amplitude in the combined limit, the Baker-Campbell-Hausdorff formula is employed to expand the matrix product $HH'u'u^{-1}$.
The Baker-Campbell-Hausdorff formula expands the product of two matrix exponentials as
\be
\begin{split}
e^X e^Y =& \exp{\left(X + Y + \frac{1}{2}[X,Y]\right.}\\
 +& \left.\frac{1}{12}([X,[X,Y]] + [Y,[Y,X]]) + \cdots \right).
\end{split}
\ee
Using this formula, the matrix $HH'$ can be expanded as
\be
\begin{split}
HH' =& \exp{(v^a \lambda^a)} \exp{((v^a+\delta v^a) \lambda^a)} \\
=&
\exp{
    \left(
        2
        v^a\lambda^a
        +
        \delta v^a \lambda^a
    \right.
}\\
    &+
    \left.
        \ii 
        v^c 
        \delta v^b
        f^{cba}\lambda^a
        +
        \cdots
    \right),
\end{split}
\ee
and $u'u^{-1}$ can be expanded as
\be
\begin{split}
u'u^{-1} =& \exp{( i (w^a+\delta w^a) \lambda^a)} \exp{(- i w^a \lambda^a)} \\
=&
\exp{
    \left(
        \ii \delta w^a\lambda^a
        +
        \ii 
        \delta w^c
        w^b
        f^{cba}\lambda^a
    \right.
}\\
    &+
    \left.
        \frac{2}{3}
        \ii 
        \delta w^b
        w^a 
        w^c
        f^{cbd}f^{adk}\lambda^{k}
        +
        \cdots
    \right).
\end{split}
\ee
For convenience, assume $\delta v^a$ and $\delta w^a$ are of the same order. Then, the first-order expansion of the matrix $HH'u'u^{-1}$ is given by \eqref{eq:HHuuexpansion}. 
As a result, the magnitude parameter $\langle \tilde{v} \rangle^2 = \operatorname{Tr}(W^2)/2$ is given by 
\be
\label{eq:TrW2}
\begin{split}
\operatorname{Tr}(W^2)/2 &=
\left(
        4v^av^a
        +
        4\delta v^a 
        v^a
        +
        4\ii 
        \delta w^a
        v^a
\right.\\
        &+
        \left.
        4\ii 
        w^b
        v^a
        \delta w^c
        f^{cba}
        \right.
\\
        &+
        \left.
        \frac{8}{3}
        \ii
        w^a 
        w^c
        v^{k}
        \delta w^b
        f^{cbd}f^{adk}
        +
        \cdots
\right)
\end{split}
\ee
where $\delta$ represents the order of $\delta v^a$ and $\delta w^a$.
Define $\delta v^a(\tau)$ and $\delta w^a(\tau)$ as $\tau$-dependent functions such that $\delta v^a(0) = 0$ and $\delta w^a(0) = 0$ respectively.
In the limit $t\rightarrow 0$ the leading-order of the overlap amplitude is denoted as 
\be
\begin{split}
& k^{t(0)}(g,g') \\
=&
e^{\frac{
        \langle \tilde{v} \rangle^2-4\langle v \rangle^2+4\langle v' \rangle^2
    }{
        t
    }
}
\Psi(\tilde{v}, v, v', \tilde{\eta}, \eta, \eta'),
\end{split}
\ee
where $\Psi(\tilde{v}, v, v', \tilde{\eta}, \eta, \eta')$ is $\frac{|\Phi(\langle \tilde{v} \rangle, \tilde{\eta})|^2}{\Phi(2\langle v' \rangle, 8\eta')\Phi(2\langle v \rangle, 8\eta)}$.
Under the semiclassical limit, the expansion of $k^t(g,g')$ around $\tau=0$ is given by
\be
\begin{split}
&k^{t(0)}(g,g')\\
=&
k^{t(0)}(g,g)\\
+&
\left[\frac{1}{t}
\partial_{\tau}
(\langle \tilde{v} \rangle^2-4\langle v \rangle^2+4\langle v' \rangle^2)
\right]\bigg|_{\tau=0}
k^{t(0)}(g,g)
\dd \tau
\\
+&
\left[
e^{\frac{
        \langle \tilde{v} \rangle^2-4\langle v \rangle^2+4\langle v' \rangle^2
    }{
        t
    }
}
\partial_\tau\left(
\Psi(\tilde{v}, v, v', \tilde{\eta}, \eta, \eta')
\right)
\right]\bigg|_{\tau=0}
\dd \tau
\\
+&\mathcal{O}(\dd\tau^2)
,
\end{split}
\ee
where the second term dominates overlap amplitude up to the first-order expansion in the limit $t\rightarrow 0$.
Since $k^{t(0)}(g,g) = 1$, the overlap amplitude in the limit $t\rightarrow 0$ yields
\be
\begin{split}
&k^{t(0)}(g,g')\\
=&
\left[\frac{1}{t}
\partial_{\tau}
(\langle \tilde{v} \rangle^2-4\langle v \rangle^2+4\langle v' \rangle^2)
\right]\bigg|_{\tau=0}
\dd \tau
+\mathcal{O}(\dd\tau^2).
\end{split}
\ee
By \eqref{eq:TrW2}, the term $\partial_\tau(\langle \tilde{v} \rangle^2-4\langle v \rangle^2+4\langle v' \rangle^2)$ reads:
\be
\begin{split}
    &\partial_{\tau}(\langle \tilde{v} \rangle^2-4\langle v \rangle^2+4\langle v+\delta v \rangle^2)\\
    =&
    8\ii 
    v^a
    \partial_\tau\delta w^a
    +
    8\ii 
    w^b
    v^a
    f^{cba}
    \partial_\tau\delta w^c
    \\
    +&
    \frac{16}{3}
    \ii
    w^a 
    w^c
    v^{k}
    f^{cbd}f^{adk}
    \partial_\tau\delta w^b
    +
    \cdots\\
    =&
    8\ii 
    v^a
    \partial_\tau\delta w^c
    \left(
    \delta^{ac}
    +
    w^b
    f^{cba}
    +
    \frac{2}{3}
    w^k 
    w^b
    f^{bcd}f^{kda}
    \right)
    \\
    +&
    \cdots\\
    =&
    K^{ac}
    v^a
    \partial_\tau\delta w^c+\cdots,
\end{split}
\ee
where $K^{ac}$ is defined as
\be
K^{ac} :=
8\ii
\left(
    \delta^{ac}
    +
    w^b
    f^{cba}
    +
    \frac{2}{3}
    w^k 
    w^b
    f^{bcd}f^{kda}
\right).
\ee
Thus, the leading order expression of the overlap amplitude in the combined limit is given by
\be
k^{t(0)}(g,g')
=
\frac{1}{t}
K^{ac}
v^a
\partial_\tau\delta w^c
\dd \tau
+\mathcal{O}(\dd\tau^2).
\label{eq:overlap_amplitude_final}
\ee
\section{Conclusion and Discussion}
In this work, we have successfully generalized the construction of diffeomorphism-covariant gauge field coherent states from the well-established $SU(2)$ case to the physically crucial $SU(3)$ gauge group. This represents a significant step toward developing a semiclassical framework for non-Abelian Yang-Mills theories within the background-independent context of loop quantum gravity.

The primary achievement of this paper is the explicit construction of $SU(3)$ coherent states using the heat kernel complexifier method. We detailed the necessary ingredients for this construction, including the dimensions of irreducible representation \eqref{eq:dimension}, eigenvalues of the Laplacian \eqref{eq:laplacian} for $SU(3)$ irreducible representations labeled by Dynkin indices $(p,q)$, and the corresponding character formula \eqref{eq:character}. The resulting coherent state wave function on a single edge is given by the sum over representations in \eqref{eq:SU3_coherent_state}.

The core of our analysis was dedicated to establishing the semiclassical properties of these states. Through a detailed asymptotic analysis in the limit $t\rightarrow 0 $, which involved extending the summation over representations via the Poisson summation formula and evaluating the resulting integrals via the method of steepest descent, we derived the leading order expression for the coherent state \eqref{eq:coherent_final_asymptotics}. This expression was further refined by solving the matrix exponential problem for $SU(3)$, allowing us to express the state in terms of the magnitude parameter $\langle v \rangle$ and the determinant parameter $\eta$ of the complexified group element $gh^{-1}$, as shown in \eqref{eq:coherent_semi_final_1}.

We numerically verified the essential peakedness property. Using MCMC methods, we demonstrated that the probability density $p^t_g(h)$ defined in \eqref{eq:peakedness} is sharply peaked at the classical configuration $h=u$, where $g=Hu$ is the polar decomposition, as $t\rightarrow 0 $. The results, visualized in Fig. \ref{fig:mcmc1}, confirm that the coherent states are sharply peaked in the configuration space.

Furthermore, we analyzed the overlap function between two coherent states. Our numerical simulations, presented in Fig. \ref{fig:mcmc_overlap}, confirm that the overlap $i^t(g,g')$ decays rapidly when $g$ and $g'$ are distinct in the classical phase space, demonstrating the desired resolution of the identity. Crucially, we also computed the leading order behavior of the overlap amplitude $k^t(g,g')$ in the combined limit $t\rightarrow 0$ and $g\rightarrow g'$, which is essential for the saddle-point approximation in the path integral formulation. The result, given in \eqref{eq:overlap_amplitude_final}, shows a Gaussian decay controlled by a kernel $K^{ac}$ that encodes the semiclassical behavior of the overlap amplitude.

The successful construction and analysis of $SU(3)$ coherent states open several avenues for future research. The most immediate application is the formulation of a coherent state path integral for $SU(3)$ gauge theories coupled to gravity. With the overlap amplitude now defined, one can proceed to construct the path integral on a fixed graph following the methodology of \cite{PhysRevD.102.024083,Han:2020uhb,Han:2021cwb,Zhang:2022vsl}. The saddle-point approximation of this path integral is expected to yield effective equations that incorporate quantum gravity corrections to the classical Yang-Mills dynamics. This provides a concrete pathway to explore phenomenological implications of quantum gravity for the strong interaction in a background-independent setting.

However, several challenges and extensions remain:
\begin{itemize}
    \item \textbf{Extension to larger gauge groups.} While $SU(3)$ is of primary physical interest, extending the construction to larger gauge groups such as $SU(N)$ for $N>3$ would be valuable for exploring grand unified theories within the loop quantum gravity framework.
    \item \textbf{Ehrenfest theorem.} A rigorous proof of the Ehrenfest theorem for $SU(3)$ coherent states is necessary to confirm that expectation values of quantum operators reproduce classical equations of motion in the semiclassical limit.
    \item \textbf{Symplectic structure in the phase space.} The overlap amplitude suppose to provides a insight into the symplectic structure of the classical phase space. In \cite{PhysRevD.102.024083}, this relation is shown for $SU(2)$ group. Extending this analysis to $SU(3)$ would deepen our understanding of the classical-quantum correspondence in non-Abelian gauge theories.
\end{itemize}
In conclusion, this work lays the groundwork for a semiclassical treatment of $SU(3)$ gauge theories within loop quantum gravity, opening new directions for both theoretical exploration and potential phenomenological applications.

\acknowledgments
Zichang Huang thanks Hongguang Liu's helpful discussions. This work is supported by the starting grant of the University of Shanghai for Science and Technology.
\subsection*{DATA AVAIlLABILITY}
The data that support the findings of this article are openly available in \cite{gitcode}.
\bibliographystyle{unsrt}
\bibliography{graviton.bib}
\newpage
\onecolumngrid
\appendix

\section{Appendix: Calculation Details} 
\be
\label{eq:summand_final}
\begin{split}
&\sum_{i=1}^6Term_i^{(0)}\\
=&
\sum_{K_1}\sum_{K_2}
-8i\sqrt{3} e^{-\frac{2(4K1^2\pi^2 + 4K2^2\pi^2 + \theta_1^2 + 2K2\pi(\theta_1 - \theta_2) + \theta_1\theta_2 + \theta_2^2 - 2K1\pi(2K2\pi + 2\theta_1 + \theta_2))}{\pi t}} \\
& \quad \times (2K1\pi + 2K2\pi - \theta_1 - 2\theta_2)(-2K1\pi + 4K2\pi + \theta_1 - \theta_2)(-4K1\pi + 2K2\pi + 2\theta_1 + \theta_2) \\
& + 8i\sqrt{3} e^{-\frac{2((2K1\pi + \theta_1)^2 - (2K1\pi + \theta_1)(2K2\pi + \theta_1 + \theta_2) + (2K2\pi + \theta_1 + \theta_2)^2)}{\pi t}} \\
& \quad \times (2K1\pi - 4K2\pi - \theta_1 - 2\theta_2)(4K1\pi - 2K2\pi + \theta_1 - \theta_2)(2K1\pi + 2K2\pi + 2\theta_1 + \theta_2) \\
& + 8i\sqrt{3} e^{-\frac{2((2K1\pi + \theta_1)^2 - (2K1\pi + \theta_1)(2K2\pi - \theta_2) + (-2K2\pi + \theta_2)^2)}{\pi t}} \\
& \quad \times \pi t (2K1\pi + 2K2\pi + \theta_1 - \theta_2)(4K1\pi - 2K2\pi + 2\theta_1 + \theta_2)(2K1\pi - 4K2\pi + \theta_1 + 2\theta_2) \\
& - 8i\sqrt{3} e^{-\frac{2((-2K2\pi + \theta_1)^2 - (2K2\pi - \theta_1)(2K1\pi + \theta_2) + (2K1\pi + \theta_2)^2)}{\pi t}} \\
& \quad \times \pi t (2K1\pi + 2K2\pi - \theta_1 + \theta_2)(2K1\pi - 4K2\pi + 2\theta_1 + \theta_2)(4K1\pi - 2K2\pi + \theta_1 + 2\theta_2) \\
& + 8i\sqrt{3} e^{-\frac{2(4K1^2\pi^2 + 4K2^2\pi^2 + \theta_1^2 + \theta_1\theta_2 + \theta_2^2 + 2K2\pi(-\theta_1 + \theta_2) - 2K1\pi(2K2\pi + \theta_1 + 2\theta_2))}{\pi t}} \\
& \quad \times (2K1\pi + 2K2\pi - 2\theta_1 - \theta_2)(-2K1\pi + 4K2\pi - \theta_1 + \theta_2)(-4K1\pi + 2K2\pi + \theta_1 + 2\theta_2) \\
& + 8i\sqrt{3} e^{-\frac{2((2K1\pi + \theta_2)^2 - (2K1\pi + \theta_2)(2K2\pi + \theta_1 + \theta_2) + (2K2\pi + \theta_1 + \theta_2)^2)}{\pi t}} \\
& \quad \times (2K1\pi - 4K2\pi - 2\theta_1 - \theta_2)(4K1\pi - 2K2\pi - \theta_1 + \theta_2)(2K1\pi + 2K2\pi + \theta_1 + 2\theta_2)
\end{split}
\ee

Computation of the norm of the coherent state:
\be
\label{eq:norm_calculation}
\begin{split}
\|\psi^t_g\|^2
= &
\int d\mu(h) 
    \sum_\pi \sum_{\pi^\prime} \\
    \times&
    d_\pi e^{-\frac{t}{2} \lambda_\pi} \overline{\pi^{\gamma\alpha}\left(H uh^{-1}\right)}\delta^{\alpha\gamma}\\
    \times&
    d_{\pi^\prime} e^{-\frac{t}{2} \lambda_{\pi^\prime}} {\pi^\prime}^{\zeta\beta}\left(H uh^{-1}\right)\delta^{\beta\zeta}
\\
= &
\int d\mu(uh^{-1}) 
    \sum_\pi \sum_{\pi^\prime} \\
    \times&
    d_\pi e^{-\frac{t}{2} \lambda_\pi} 
    \overline{\pi^{\gamma\sigma}\left(H\right)}
    \overline{\pi^{\sigma\alpha}\left( uh^{-1}\right)}
    \delta^{\alpha\gamma}\\
    \times&
    d_{\pi^\prime} e^{-\frac{t}{2} \lambda_{\pi^\prime}} {\pi^\prime}^{\zeta\phi}\left(H \right)
    {\pi^\prime}^{\phi\beta}\left(uh^{-1}\right)
    \delta^{\beta\zeta}
\\
= &
    \sum_\pi  d_\pi e^{-t \lambda_\pi} 
    \overline{\pi^{\gamma\sigma}\left(H\right)}
    \delta^{\alpha\gamma}
    \delta^{\sigma\phi}
    {\pi}^{\zeta\phi}\left(H \right)
    \delta^{\alpha\beta}
    \delta^{\beta\zeta}
\\
= &
    \sum_\pi  d_\pi e^{-t \lambda_\pi} 
    \overline{\pi^{\alpha\sigma}\left(H\right)}
    {\pi}^{\alpha\phi}\left(H \right)
    \delta^{\sigma\phi}
\\
= &
    \sum_\pi  d_\pi e^{-t \lambda_\pi} 
    \pi^{\sigma\alpha}\left(H^\dagger\right)
    {\pi}^{\alpha\phi}\left(H \right)
    \delta^{\sigma\phi}
\\
= &
    \sum_\pi  d_\pi e^{-t \lambda_\pi} 
    \operatorname{Tr}(H^2),
\\
\end{split}
\ee
where in the second equality we have used the property of representation matrices $\pi(gh) = \pi(g)\pi(h)$, in the third equality we have used the orthogonality relation \eqref{eq:orthogonality}, and in the last equality we have used the fact that $H$ is a Hermitian matrix.

\be
\label{eq:norm_calculation_2}
\begin{split}
\langle\psi^t_{g}|\psi^t_{g'}\rangle
= &
\int d\mu(h) 
    \sum_\pi \sum_{\pi^\prime} \\
    \times&
    d_\pi e^{-\frac{t}{2} \lambda_\pi} \overline{\pi^{\gamma\alpha}\left(H uh^{-1}\right)}\delta^{\alpha\gamma}\\
    \times&
    d_{\pi^\prime} e^{-\frac{t}{2} \lambda_{\pi^\prime}} {\pi^\prime}^{\zeta\beta}\left(H' u'h^{-1}\right)\delta^{\beta\zeta}
\\
= &
\int d\mu(h^{-1}) 
    \sum_\pi \sum_{\pi^\prime} \\
    \times&
    d_\pi e^{-\frac{t}{2} \lambda_\pi} 
    \overline{\pi^{\gamma\sigma}\left(Hu\right)}
    \overline{\pi^{\sigma\alpha}\left( h^{-1}\right)}
    \delta^{\alpha\gamma}\\
    \times&
    d_{\pi^\prime} e^{-\frac{t}{2} \lambda_{\pi^\prime}} {\pi^\prime}^{\zeta\phi}\left(H' u' \right)
    {\pi^\prime}^{\phi\beta}\left(h^{-1}\right)
    \delta^{\beta\zeta}
\\
= &
    \sum_\pi  d_\pi e^{-t \lambda_\pi} 
    \overline{\pi^{\gamma\sigma}\left(Hu\right)}
    \delta^{\alpha\gamma}
    \delta^{\sigma\phi}
    {\pi}^{\zeta\phi}\left(H'u' \right)
    \delta^{\alpha\beta}
    \delta^{\beta\zeta}
\\
= &
    \sum_\pi  d_\pi e^{-t \lambda_\pi} 
    \overline{\pi^{\alpha\sigma}\left(Hu\right)}
    {\pi}^{\alpha\phi}\left(H'u' \right)
    \delta^{\sigma\phi}
\\
= &
    \sum_\pi  d_\pi e^{-t \lambda_\pi}
    \pi^{\sigma\beta}\left(u^\dagger\right) 
    \pi^{\beta\alpha}\left(H^\dagger\right)
    {\pi}^{\alpha\phi}\left(H' \right)
    {\pi}^{\phi\kappa}\left(u' \right)
    \delta^{\sigma\kappa}
\\
= &
    \sum_\pi  d_\pi e^{-t \lambda_\pi} 
    \operatorname{Tr}(HH'u'u^{-1}),
\\
\end{split}
\ee

First order expansion of $HH'u'u^{-1}$:
\be
\label{eq:HHuuexpansion}
\begin{split}
HH'u'u^{-1} =& 
\exp{
    \left(
        2
        v^a\lambda^a
        +
        \delta v^a \lambda^a
        +
        \ii 
        v^c 
        \delta v^b
        f^{cba}\lambda^a
        +
        \mathcal{O}((\delta v)^2)
    \right)
}\\
\times&
\exp{
    \left(
        \ii \delta w^a\lambda^a
        +
        \ii 
        \delta w^c
        w^b
        f^{cba}\lambda^a
        +
        \frac{2}{3}
        \ii 
        \delta w^b
        w^a 
        w^c
        f^{cbd}f^{adk}\lambda^{k}
        +
        \mathcal{O}((\delta w)^2)
    \right)
}    
\\
=&
\exp
    \left(
        2
        v^a\lambda^a
        +
        \delta v^a \lambda^a
        +
        \ii 
        v^c 
        \delta v^b
        f^{cba}\lambda^a
        +
        \ii \delta w^a\lambda^a
        +
        \ii 
        \delta w^c
        w^b
        f^{cba}\lambda^a
    \right.\\
    &+
    \left.
        \frac{2}{3}
        \ii 
        \delta w^b
        w^a 
        w^c
        f^{cbd}f^{adk}\lambda^{k}
        -
        2 v^b\delta w^af^{bak}\lambda^k
        -
        2
        v^d 
        \delta w^c
        w^b
        f^{dak}f^{cba}\lambda^k
    \right.\\
    &-
    \left.
        \frac{4}{3}
        v^g 
        \delta w^b
        w^a 
        w^c
        f^{gkl}f^{cbd}f^{adk}\lambda^{l}
        -
        \frac{4}{3}
        \ii
        v^g
        v^b
        \delta w^a
        f^{bak}
        f^{gkl}
        \lambda^l
    \right.\\
    &-
    \left.
        \frac{4}{3}
        \ii
        v^g
        v^d
        w^b
        \delta w^c
        f^{cba}f^{dak}
        f^{gkl}
        \lambda^l
        -
        \frac{8}{9}
        \ii
        v^h
        v^g
        w^a 
        w^c
        \delta w^b
        f^{cbd}f^{adk}f^{gkl}
        f^{hlm}
        \lambda^m
        +
        \cdots
    \right)
\end{split}
\ee

\section{MCMC Sampling Results}
\begin{figure}[htbp]
    \centering
    \includegraphics[width=0.8\textwidth]{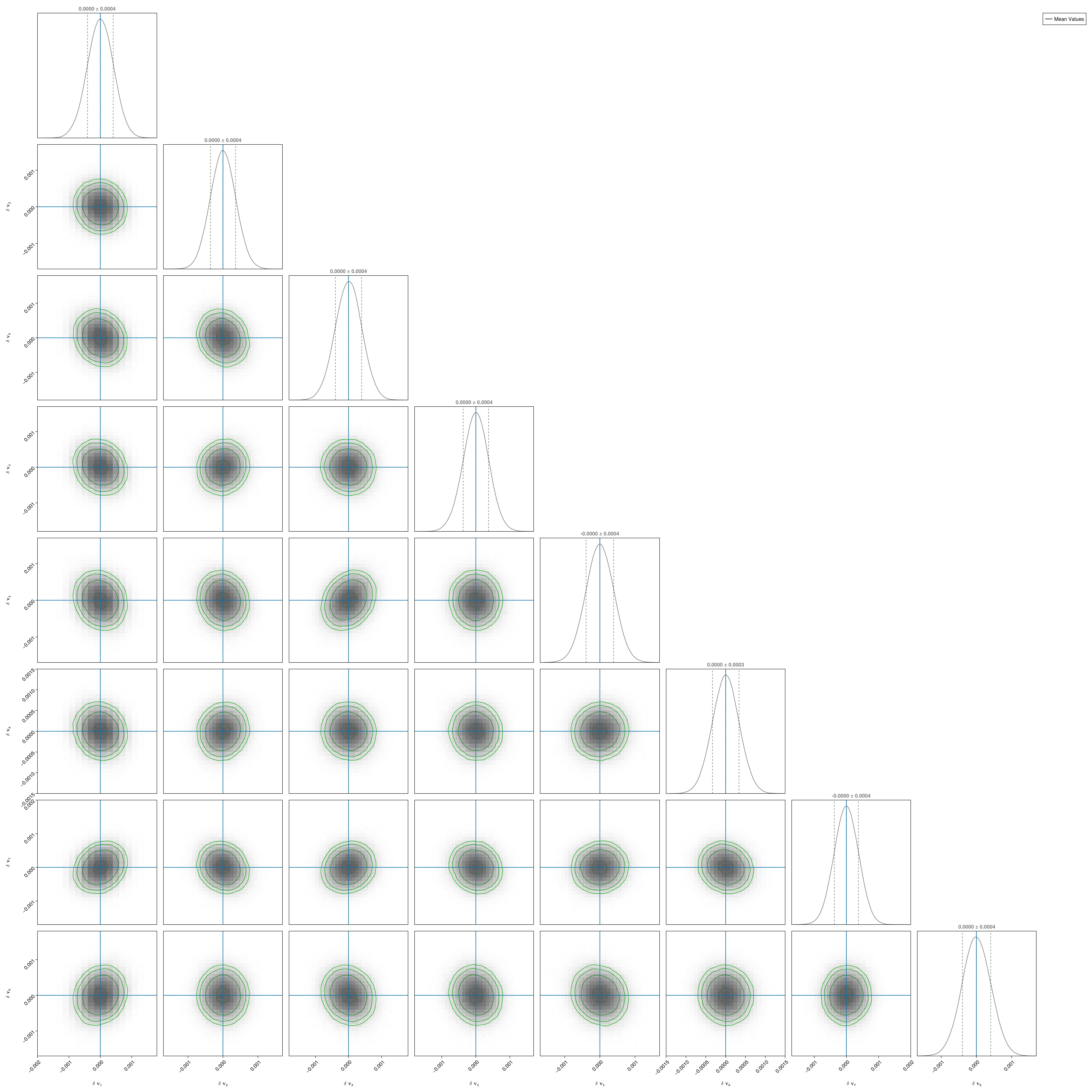}
    \caption{MCMC sampling results of the peak position of the coherent state for a randomly chosen $H$ and $u$. The diagonal subplots show the marginal distribution of each $\delta v^a$ for $a = 1,2,\cdots, 8$.  The lower triangle subplots show the covariance between any pair of $\delta v^a$. In each of these subplots, the 2-dimensional histogram of the sampled points is shown by the grey scale, the blue cross resepents the mean value of the samples, and the green ellipse represents the contour of the standard deviation at one sigma, two sigma, three sigma, and four sigma levels respectively. The results show that the mean values of all $\delta v^a$ parameters are close to zero, and the covariance between any pair is smaller than standard value for two sigma level. Thus, the results confirms that the probability density $p^t_g(h)$ is sharply peaked at $h=u$ in the semiclassical limit.}
    \label{fig:mcmc1}
\end{figure}

The Fig. \ref{fig:mcmc1} shows the MCMC sampling results for seeking the peakedness of the probability density function. As discussed in the main text, in the semiclassical limit, the probability density function $p^t_g(h)$ is dominated by the exponential factor $e^{\frac{Re(\langle v \rangle)^2}{t}}$. Therefore, we use MCMC method to sample $\delta v^a$ according to the Boltzmann factor $e^{\beta Re(\langle v \rangle)^2}$, where the factor $\beta = \frac{1}{t}$ plays the role of the inverse temperature in statistical mechanics. One would expect that in the \textit{Low temperature} limit $t\rightarrow 0$, the sampled $\delta v^a$ would cluster around the point where $Re(\langle v \rangle)^2$ reaches its maximum.
The sampling is performed when $v^a$ and $w^a$ are chosen as the values shown in the TABLE. \ref{tab:vw_values}. After a burn-in period of $10^5$ steps, we collected $20^5$ samples to ensure statistical reliability.
The result samples are highly concentrated around the point $\delta v^a = 0$ for all $a=1,2,\cdots,8$, confirming the desired peakedness property of the coherent state.

\begin{table}[htbp]
\label{tab:vw_values}
\centering
\caption{The values of $v^a$ and $w^a$ used in the MCMC sampling shown in Fig. \ref{fig:mcmc1}.}
\begin{tabular}{|c|c|c|c|c|c|c|c|c|}
\hline
 &$a=1$ & $a=2$ & $a=3$ & $a=4$ & $a=5$ & $a=6$ & $a=7$ & $a=8$ \\
\hline
$w^a$ & 0.888812 & 0.295318 & 0.183671 & 0.467503 & 0.977395 & 0.502118 & 0.484956 & 0.812834 \\
\hline
$v^a$ & 0.693846 & 0.828026 & 0.264855 & 0.414134 & 0.230485 & 0.888528 & 0.345796 & 0.635267 \\
\hline
\end{tabular}
\end{table}

\begin{figure}[htbp]
    \centering
    \includegraphics[width=0.9\textwidth]{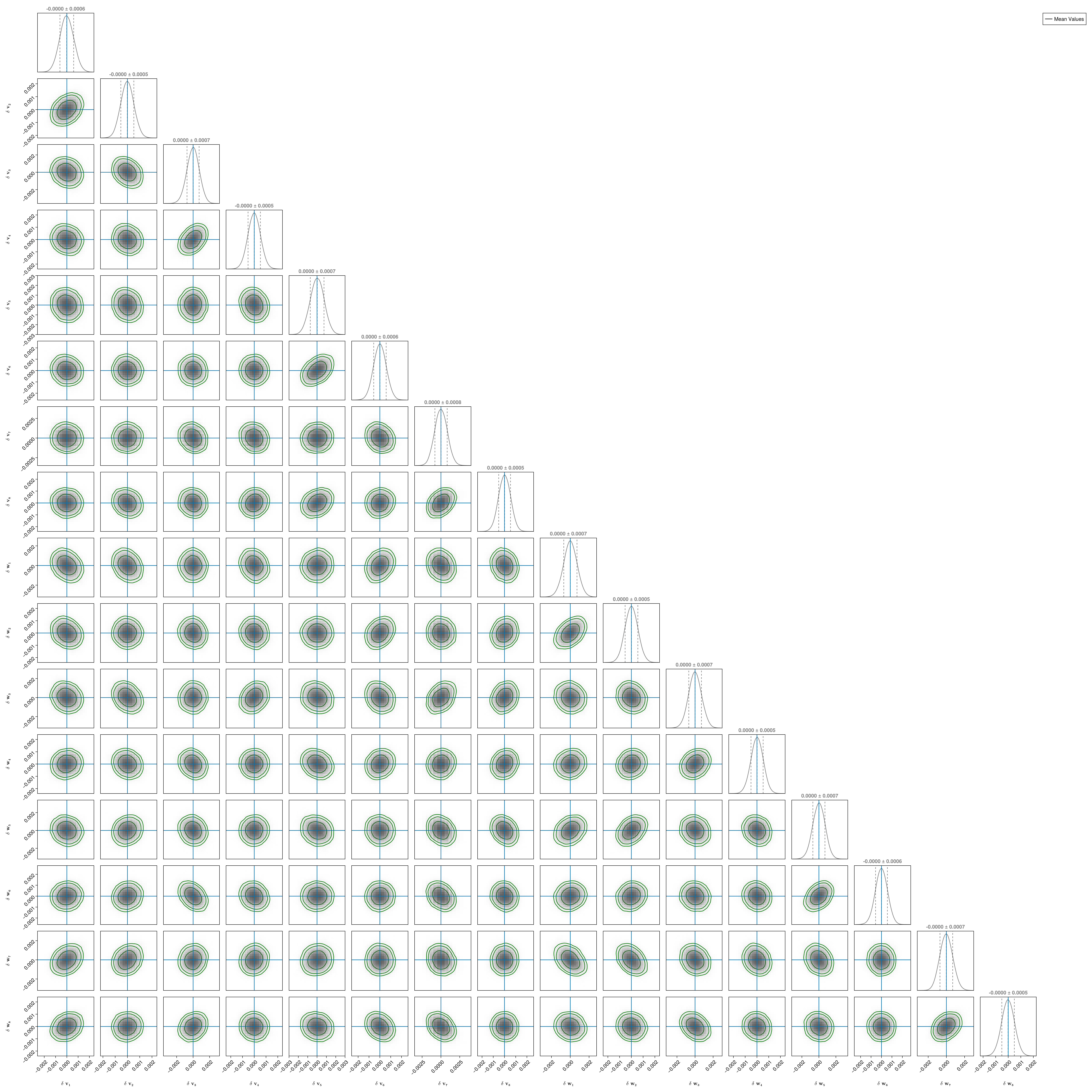}
    \caption{MCMC sampling results of the peak position of the overlap function. The diagonal subplots show the marginal distribution of each $\delta v^a$ and $\delta w^a$ for $a = 1,2,\cdots, 8$.  The lower triangle subplots show the covariance between any pair of $\delta v^a$ and $\delta w^a$. In each of these subplots, the 2-dimensional histogram of the sampled points is shown by the grey scale, the blue cross locates the mean value of the samples, and the green ellipse represents the contour of the standard deviation at one sigma, two sigma, three sigma, and four sigma levels respectively. The results show that the mean values of all $\delta v^a$ and $\delta w^a$ parameters are close to zero, and the covariance between any pair is smaller than standard value for two sigma level. Thus, the results confirms that the overlap function is sharply peaked at $g=g'$ in the semiclassical limit.}
    \label{fig:mcmc_overlap}
\end{figure}

The Fig. \ref{fig:mcmc_overlap} shows the results of MCMC sampling of the peak of the overlap function. Similar to the previous case, we use MCMC method to sample the parameters $\delta v^a$ and $\delta w^a$ according to the Boltzmann factor $e^{\beta (Re(\langle \tilde{v} \rangle)^2-4\langle v \rangle^2+4\langle v' \rangle^2)}$, where $\beta = \frac{1}{t}$ is the inverse temperature. In the \textit{Low temperature} limit $t\rightarrow 0$, the sampled $\delta v^a$ and $\delta w^a$ would cluster around the point where $Re(\langle \tilde{v} \rangle)^2-4\langle v \rangle^2+4\langle v' \rangle^2$ reaches its maximum.
In this sampling, the values of $v^a$, and $w^a$ are also chosen as those shown in the TABLE. \ref{tab:vw_values}. The parameter $\beta$ is set to be $10^6$. After a burn-in period of $10^5$ steps, we collected $20^5$ samples to ensure statistical reliability.
The result samples are highly concentrated around the point $\delta v^a = 0$ and $\delta w^a = 0$ for all $a=1,2,\cdots,8$, confirming the desired peakedness property of the overlap function.

\end{document}